%% file: Fault_Inception_Detection.tex
\documentclass[conference]{IEEEtran}
\IEEEoverridecommandlockouts

\usepackage{cite}
\usepackage{amsmath,amssymb,amsfonts}
\usepackage{algorithmic}
\usepackage{graphicx}
\usepackage{tabularx}
\usepackage{textcomp}
\usepackage{xcolor}
\usepackage{acronym}
\usepackage{siunitx}
\usepackage{booktabs}
\usepackage{threeparttable}
\usepackage{url}
\usepackage{hyperref}

\def\BibTeX{{\rm B\kern-.05em{\sc i\kern-.025em b}\kern-.08em
    T\kern-.1667em\lower.7ex\hbox{E}\kern-.125emX}}

\begin{document}
\input{acronyms}

\title{Fault Inception Detection in Real-World Disturbance Data for Power System Protection%
\thanks{\textcopyright{} 2026 IEEE. Personal use of this material is permitted.  Permission from IEEE must be obtained for all other uses, in any current or future media, including reprinting/republishing this material for advertising or promotional purposes, creating new collective works, for resale or redistribution to servers or lists, or reuse of any copyrighted component of this work in other works.}
}

\author{
\IEEEauthorblockN{
Julian Oelhaf\textsuperscript{1}\textsuperscript{$\dagger$}\textsuperscript{*},
Mehran Pashaei\textsuperscript{1}\textsuperscript{$\dagger$},
Paula Andrea Pérez-Toro\textsuperscript{1},
Georg Kordowich\textsuperscript{2},
Christian Bergler\textsuperscript{3}\\
Andreas Maier\textsuperscript{1},
Johann J\"ager\textsuperscript{2},
Siming Bayer\textsuperscript{1}
}

\IEEEauthorblockA{
\textit{\textsuperscript{1}Pattern Recognition Lab, Friedrich-Alexander-Universit\"at Erlangen-N\"urnberg} \\
\textit{\textsuperscript{2}Institute of Electrical Energy Systems, Friedrich-Alexander-Universit\"at Erlangen-N\"urnberg} \\
\textit{\textsuperscript{3}Department of Electrical Engineering, Media and Computer Science, Ostbayerische Technische Hochschule Amberg-Weiden} \\
{\footnotesize \textsuperscript{$\dagger$}These authors contributed equally.}\\
{\textsuperscript{*}Corresponding author: julian.oelhaf@fau.de}
}
}

\maketitle

\begin{abstract}
Large collections of real-world disturbance recordings are increasingly available in transmission networks, but their value for power system protection and automated disturbance analysis is limited by the absence of precise event-onset annotations. In practice, field-recorded voltage and current waveforms contain switching operations, transformer energization, resonance, saturation, and other non-ideal effects that can obscure or mimic genuine fault signatures, making reliable fault inception detection difficult. This paper presents an training-free framework for fault inception detection in real-world transmission disturbance data. The method combines protection-domain indicators, robust median/MAD-based normalization, a low-latency transient path, and persistence-aware fusion and veto logic to distinguish fault-consistent disturbances from non-fault transients. We apply the framework to 12{,}053 transmission-level recordings from the publicly available RTE database and further assess detector performance on a manually reviewed subset of 300 events. On the reviewed subset, the detector achieves 96.6\% recall, 79.2\% precision, and a median timing error of 4.2\,ms for matched detections. These results indicate that the proposed approach can support protection-oriented disturbance screening, relay and post-event analysis, and the creation of timestamp annotations for downstream data-driven monitoring tasks.
\end{abstract}


\section{Introduction}

Large collections of disturbance recordings are increasingly available in transmission networks, typically captured by high-resolution \acp{dfr} as synchronized voltage and current waveforms during grid events~\cite{phadke_computer_2009,perez_guide_2010}. These data support protection analysis, relay benchmarking, post-event investigation, and data-driven monitoring~\cite{kezunovic_smart_2011}. However, many recorded events correspond to switching actions or operational procedures rather than actual faults, making manual analysis difficult to scale~\cite{moreto_using_2015}. 

Automatic estimation of the fault inception time $t_0$ in field recordings remains challenging. Switching operations, transformer energization, resonance, saturation, and measurement noise can mimic or obscure true fault signatures~\cite{wang_analysis_2008}. Conventional normalization is vulnerable to statistical masking, where large disturbances inflate variance estimates and suppress anomaly scores near inception~\cite{leys_detecting_2013,rousseeuw_robust_2011}. These factors make manual annotation costly and limit scalability across large disturbance datasets~\cite{boyd_learning_2024}. While recent work highlights the potential of large disturbance archives, they remain largely unlabeled in practice~\cite{oelhaf_unsupervised_2025}.

Fault detection and disturbance analysis are widely studied using protection-domain methods, signal processing, and data-driven approaches. Classical techniques rely on relay principles and time--frequency analysis, such as wavelet-based transient detection~\cite{zhang_transmission_2007}. Rule-based and expert systems enable interpretable classification and diagnostic support for disturbance records~\cite{boyd_learning_2024,hossack_multiagent_2003}. Machine learning and sequential detection methods have demonstrated strong performance in fault detection and classification~\cite{rodrigues_power_2023,mozaffari_real-time_2022}. However, these approaches typically assume labeled data, simulated conditions, or predefined fault intervals. While they detect or classify disturbances, explicit estimation of the fault inception time $t_0$ is rarely addressed as a standalone problem, particularly in real, unlabeled disturbance recordings.

To address this gap, this paper proposes a physics-guided framework for fault inception detection in transmission disturbance data. The method combines complementary protection-domain indicators with robust normalization and rule-based fusion to localize disturbance onset in real-world recordings. In contrast to prior work that focuses on disturbance detection or classification, the proposed approach explicitly targets fault inception localization in unlabeled field recordings.

The main contributions are as follows:
\begin{itemize}
    \item A physics-guided framework for fault inception detection that combines cycle-synchronous protection indicators with a low-latency transient path and rule-based fusion.
    \item A robust normalization and scoring approach that mitigates statistical masking effects in disturbance recordings.
    \item Large-scale validation on 12{,}053 recordings from the RTE database~\cite{presvots_database_2024}, including manual review of 300 events, demonstrating applicability to disturbance screening, onset annotation, and protection analysis workflows.
\end{itemize}

The generated annotations and reference implementation are publicly available on GitHub to support reproducibility and further research: \href{https://github.com/pashaeimehran/fault-inception-detection}{https://github.com/pashaeimehran/fault-inception-detection}.

\section{Framework for Fault Inception Detection}

The framework is designed for fault inception detection in real-world transmission disturbance data. The detector targets field recordings, where genuine faults must be distinguished from non-fault transients and measurement artifacts. Rather than relying on supervised or black-box models, it combines physically interpretable protection-domain indicators with robust statistical normalization and rule-based fusion.

\begin{figure}[!t]
\centering
\includegraphics[width=0.96\columnwidth]{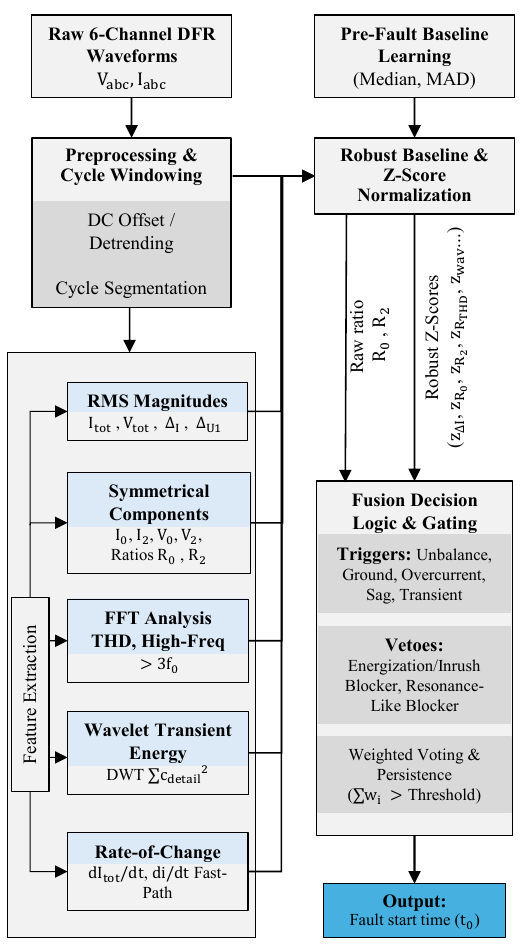}
\caption{Fusion-based fault inception detection pipeline. The method combines preprocessing, cycle-synchronous feature extraction, robust normalization, and physics-guided fusion with persistence and veto logic to estimate $t_0$.}
\label{fig:pipeline}
\end{figure}

As illustrated in Fig.~\ref{fig:pipeline}, the framework operates in a single pass. After offset removal and detrending, each event is partitioned into non-overlapping one-cycle windows of length $N_c = F_s/f_0$. For each window, features are extracted, normalized using robust baseline statistics from an initial pre-event region, and fused using persistence-aware logic with veto conditions. The output is an estimated fault inception time $t_0$.

The design follows three principles:
(i) physically interpretable indicators,
(ii) robustness to large disturbances via median/\ac{mad}-based normalization, and
(iii) persistence-aware fusion of complementary evidence.
The following subsections describe feature extraction, normalization, and the hybrid detection architecture.

\subsection{Preprocessing and Physics-Guided Feature Extraction}

Let $x[n]$ denote a waveform channel. To reduce zero-sequence artifacts caused by offsets and slow measurement drift, each channel is first de-meaned and detrended using a moving-average trend estimate. The processed signals are then partitioned into non-overlapping one-cycle windows of length $N_c = F_s / f_0 = 128$ samples, where $F_s$ is the sampling frequency and $f_0$ is the nominal system frequency. Features are extracted cycle-by-cycle to preserve synchronization with the fundamental waveform and align the detector with protection-oriented decision logic.

For each window $k$, we compute a set of physically interpretable features capturing complementary fault signatures including magnitude changes, sequence imbalance, distortion, and transient activity. First, RMS magnitudes are evaluated for each phase voltage and current and aggregated as
\begin{equation*}
\begin{aligned}
I_{\mathrm{tot}}[k] &= \sqrt{\sum_{p\in\{a,b,c\}} I_{\mathrm{rms},p}^2[k]},\\
V_{\mathrm{tot}}[k] &= \sqrt{\sum_{p\in\{a,b,c\}} V_{\mathrm{rms},p}^2[k]}.
\end{aligned}
\end{equation*}
Relative changes with respect to pre-event baseline estimates are then formed as
\begin{equation*}
\Delta I[k] = \frac{I_{\mathrm{tot}}[k]}{\widehat{I}_{\mathrm{base}}},\qquad
\Delta U_1[k] = \frac{|V_1[k]|}{\widehat{V}_{1,\mathrm{base}}},
\end{equation*}
where $V_1[k]$ denotes the positive-sequence voltage magnitude.

To capture fault-relevant imbalance, we compute Fortescue symmetrical components~\cite{phadke_computer_2009}. For each one-cycle window, the three-phase time-domain samples are transformed using the Fortescue matrix, and RMS magnitudes of the resulting zero-, positive-, and negative-sequence sample streams are used to obtain $(I_0[k], I_1[k], I_2[k])$ and $(V_0[k], V_1[k], V_2[k])$. Since sequence ratios become unstable under light-load or de-energized conditions, the current normalization is stabilized using an adaptive floor derived from the pre-event positive-sequence current:
\begin{align}
I_{1,\mathrm{floor}} &= \max\!\left(0.1\,|\tilde I_{1,\mathrm{pre}}|,\;10^{-2}\right),\\
I_{1,\mathrm{safe}}[k] &= \max\!\left(I_1[k],\, I_{1,\mathrm{floor}}\right),
\end{align}
where $\tilde I_{1,\mathrm{pre}}$ is the median of $I_1[k]$ over the pre-fault windows. The normalized imbalance indicators are then defined as
\begin{equation*}
R_2[k] = \frac{I_2[k]}{I_{1,\mathrm{safe}}[k]},\qquad
R_0[k] = \frac{I_0[k]}{I_{1,\mathrm{safe}}[k]}.
\end{equation*}

To distinguish fault-consistent events from switching-like disturbances, we further compute spectral and transient features from the current windows. A real FFT is used to derive (i) a high-frequency energy ratio above the third harmonic and (ii) total harmonic distortion (THD), which has been reported as useful for separating normal, inrush, and fault conditions~\cite{raichura_2_tddifftotal_2021}. In parallel, we compute wavelet-detail energy from a discrete wavelet decomposition, following established wavelet-based protection formulations~\cite{zhang_transmission_2007}:
\begin{equation*}
E_{\mathrm{wav}}[k] = \sum_n |c_{\mathrm{detail}}[n]|^2.
\end{equation*}
Finally, we include a rate-of-change feature based on total current RMS,
\begin{equation*}
\frac{d I_{\mathrm{tot}}}{dt}[k] = \frac{I_{\mathrm{tot}}[k]-I_{\mathrm{tot}}[k-1]}{T_c},
\end{equation*}
where $T_c = 1/f_0$ is the cycle duration. Together, these features provide complementary evidence for current increase, voltage depression, imbalance, distortion, and impulsive transients at the disturbance boundary.

\begin{table}[t]
\caption{Benchmark configuration}
\label{tab:params}
\centering
\footnotesize
\setlength{\tabcolsep}{3pt}
\renewcommand{\arraystretch}{1.1}
\begin{tabular*}{\columnwidth}{@{\extracolsep{\fill}}lclc}
\toprule
\textbf{Parameter} & \textbf{Value} & \textbf{Parameter} & \textbf{Value} \\
\midrule
$K_b$                 & 12 cycles (min.\ 5) & $\tau_{R2}$                       & 0.12 \\
$K_p$                 & 4 cycles          & $\tau_{R0}$                       & 0.05 \\
$T_f$                 & 5\,ms             & $\tau_{\Delta I}$                 & 1.3 \\
$T_{\mathrm{search}}$ & 60\,ms            & $\tau_{\Delta U}^{\mathrm{core}}$ & 0.95 \\
$S_{\mathrm{high}}$   & 3.0               & $\tau_{\Delta U}^{\mathrm{gate}}$ & 0.92 \\
$S_{\mathrm{strict}}$ & 5.0               & $\tau_z$                          & 4.0 (5.0 for $z_{R0}$ gate) \\
$w_i$                 & core $(2,2,2)$; else 1 & & \\
\bottomrule
\end{tabular*}
\end{table}

\subsection{Baseline Normalization}

A key design requirement is robustness against statistical masking, where high-energy disturbances inflate conventional mean/standard-deviation estimates and suppress anomaly scores near inception. We therefore estimate feature baselines from an initial pre-event region of $K_b$ windows using robust statistics~\cite{leys_detecting_2013}. If this initial region is short or already disturbed, the baseline may be biased, since no adaptive search for a clean pre-fault segment is performed. For each feature $x[k]$, the baseline median and \ac{mad} are
\begin{equation*}
m = \mathrm{median}(x[1{:}K_b]),\qquad
\mathrm{MAD} = \mathrm{median}(|x[1{:}K_b]-m|),
\end{equation*}
and the corresponding modified Z-score is
\begin{equation*}
z[k] = \frac{x[k]-m}{1.4826\,\mathrm{MAD}+\epsilon},
\end{equation*}
where the factor $1.4826$ provides consistency with the normal scale~\cite{leys_detecting_2013}. These normalized scores enable a common thresholding scheme across features while remaining insensitive to isolated high-amplitude excursions.

\begin{figure}
\centering
\includegraphics[width=\columnwidth]{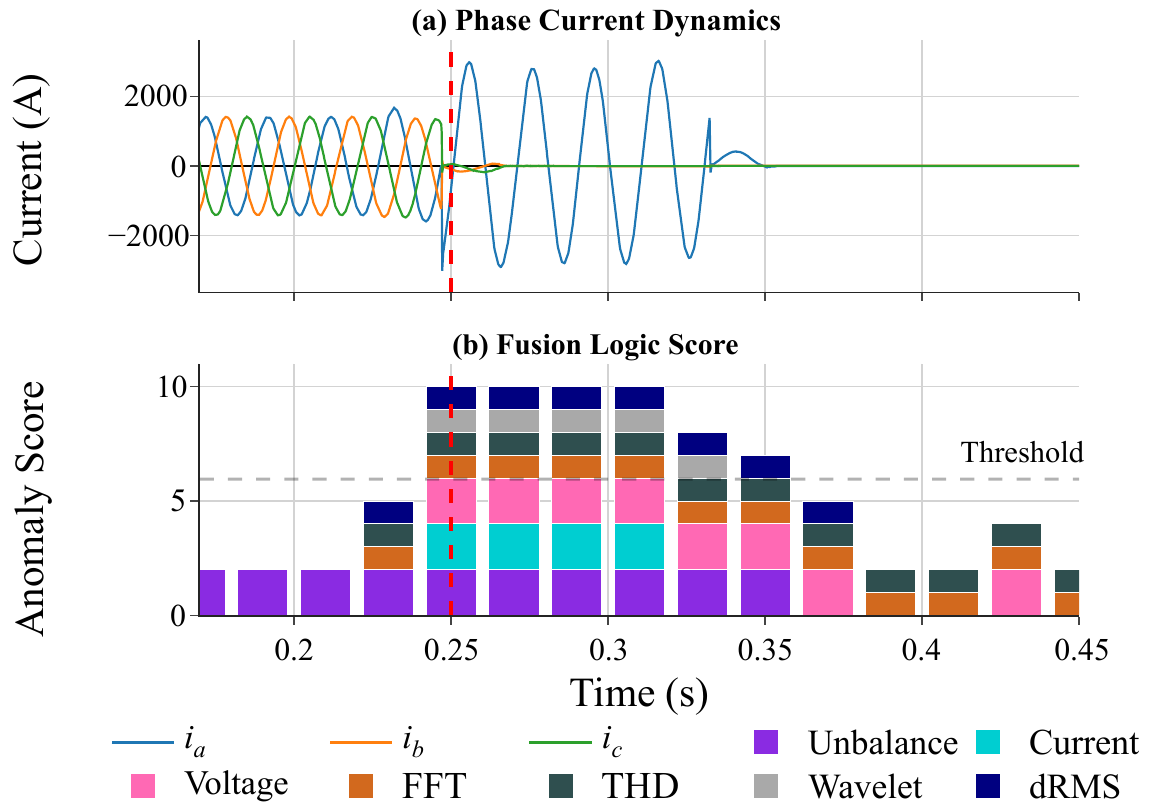}
\caption{Representative fault event in a zoomed view around the estimated inception time. (a) Currents showing the transition to an unbalanced disturbed regime. (b) Stacked fusion-score contributions. The red line marks the estimated inception time, and the gray line indicates the decision threshold.}
\label{fig:ieee_fig_event_001}
\end{figure}

\subsection{Hybrid Detection and Fusion Logic}

The detector combines two complementary decision paths. The primary path is a cycle-synchronous fusion engine operating on the windowed features described above. To reduce the inherent one-cycle latency of this pipeline, a parallel fast-path detector operates directly on the sample stream. It monitors residual zero-sequence current and voltage, $i_0[n]$ and $u_0[n]$, together with the sample-wise derivative $di_0/dt$, to detect abrupt ground-fault-like transients.

Rather than using fixed thresholds, the fast-path trigger levels are adapted to the event magnitude and early-window variability. Over a short window of duration $T_f$ (with $N_f=\lfloor T_f F_s\rfloor$ samples), a fast-path trigger is issued at the earliest sample satisfying
\begin{equation*}
|i_0[n]| > I_{0,\min}\ \ \wedge\ \ \Big(|di_0/dt| > \tau_d \ \ \vee\ \ |u_0[n]| > U_{0,\min}\Big),
\end{equation*}
with the search restricted to a bounded interval shortly after the pre-fault baseline region. This branch provides sub-cycle responsiveness for abrupt events, while the main fusion logic retains higher security for gradual or ambiguous disturbances.

The cycle-synchronous fusion stage forms Boolean trigger families from absolute thresholds and robust Z-scores. The principal trigger classes are: (i) unbalance triggers based on $R_2$; (ii) ground-fault triggers based on $R_0$; (iii) overcurrent triggers based on $\Delta I$; (iv) voltage-sag triggers based on $\Delta U_1$; and (v) distortion and transient triggers derived from FFT, THD, wavelet energy, and $dI_{\mathrm{tot}}/dt$.

Each active trigger contributes to a weighted fusion score
\begin{equation*}
S[k] = \sum_i w_i\,\mathbb{I}\{C_i[k]\},
\end{equation*}
where $C_i[k]$ is the $i$-th trigger condition and $w_i$ its weight. The weights used for this dataset are summarized in Table~\ref{tab:params}.

Decision security is enforced through persistence and veto logic. A two-level voltage criterion distinguishes healthy voltage from definite sag evidence, allowing current-based triggers to be suppressed under load-pickup-like conditions. Similarly, ground-fault candidates are blocked under resonance-like signatures by requiring magnitude supervision and consistency between zero-sequence current and voltage behavior.

A \emph{strict} decision path requires (i) a minimum score $S[k]\ge S_{\mathrm{strict}}$, (ii) a protection-relevant core conjunction of anomalies, and (iii) persistence over a confirmation window of length $K_p$. In the benchmark implementation, persistence is satisfied when core indicators remain active in at least $0.75\,K_p$ windows. In parallel, a \emph{high-sensitivity} path admits candidates with $S[k]\ge S_{\mathrm{high}}$ when they are supported by strong transient evidence and magnitude-constrained gating. This two-tier structure improves sensitivity to subtle or short-lived events while preserving security against switching, energization, and measurement artifacts. In the reference implementation, this core fusion logic is complemented by additional heuristic branches for short-lived transients and relay-cleared fault-like events to improve robustness on heterogeneous field recordings.

For each event, the detector returns an estimated fault inception time $t_0$. The detector operates in a single pass without supervised model training; all features are computed cycle-by-cycle, and the overall complexity is linear in the number of windows. Fig.~\ref{fig:ieee_fig_event_001} illustrates a representative disturbance and the corresponding fusion-score evolution around the estimated inception time.


\begin{table}[t]
\caption{RTE Dataset Summary~\cite{presvots_database_2024}}
\label{tab:dataset}
\centering
\footnotesize
\begin{tabular}{lc}
\toprule
\textbf{Item} & \textbf{Specification} \\
\midrule
Voltage level           & 90\,kV \\
Records                 & 12{,}053 \\
Channels                & 3V, 3I \\
Sampling rate $F_s$     & 6400\,Hz \\
Nominal frequency $f_0$ & 50\,Hz \\
Samples per cycle       & 128 \\
Record length           & 21{,}000 samples (3.28\,s) \\
\bottomrule
\end{tabular}
\end{table}

\section{Dataset and Evaluation}
\label{sec:dataset_eval}

We evaluate the proposed framework on the \textit{Digital Fault Recording Database} released by the French Transmission System Operator, R\'eseau de Transport d'\'Electricit\'e (RTE)~\cite{presvots_database_2024}. The dataset comprises 12{,}053 disturbance recordings from the 90\,kV transmission network. As summarized in Table~\ref{tab:dataset}, each record contains synchronized three-phase voltage and current waveforms under practical measurement conditions, including noise, saturation, and non-ideal transients. No metadata such as disturbance class, fault location, or event timestamps are provided; visual inspection suggests a mixture of fault and switching-related events.


Signals are provided as 16-bit integer samples and converted to physical units using dataset-specific scaling factors. Each channel is de-meaned and detrended prior to analysis. The signals are segmented into non-overlapping one-cycle windows; the final incomplete cycle is discarded to preserve cycle-synchronous processing.

\subsection*{Evaluation Protocol}\label{sec:eval_prot}

The objective is unsupervised detection of the fault inception time $t_0$, defined as the transition from steady-state operation to a fault-consistent disturbed regime. Since the RTE dataset does not provide localized ground-truth onset timestamps~\cite{presvots_database_2024}, direct timestamp evaluation over the full dataset is not feasible. We therefore adopt a two-level evaluation. First, performance is assessed on a manually reviewed subset ($N=300$), constructed via stratified sampling of detector branches and rejected cases, complemented by random sampling and adaptive filling. Second, aggregate statistics are reported for all 12{,}053 recordings to characterize detector behavior at scale.

Each event is reviewed using three-phase voltage and current waveforms alongside the detector estimate. Labels are \textit{correct}, \textit{slightly early}, \textit{slightly late}, \textit{wrong onset}, and \textit{no fault}. Detections within $\pm10$\,ms are \textit{correct}; deviations of 10--40\,ms are \textit{slightly early/late}. For detection metrics, \textit{correct} and \textit{slightly early/late} count as true positives, while \textit{wrong onset} and \textit{no fault} are false positives; missed fault-consistent cases are false negatives. Precision, recall, specificity, and F1-score are reported. Timing accuracy is evaluated via median absolute error and fractions within $\pm20$\,ms and $\pm40$\,ms.

As a reference, a single-feature baseline based on the maximum phase-current derivative is evaluated, where an event is declared if $\max(|d i_a/dt|, |d i_b/dt|, |d i_c/dt|)$ exceeds a robust pre-event threshold for a minimum duration.

\begin{table}[t]
\caption{Manual evaluation results on the reviewed subset.}
\label{tab:expert_overall}
\centering
\footnotesize
\setlength{\tabcolsep}{4pt}
\begin{threeparttable}
\begin{tabular*}{\columnwidth}{@{\extracolsep{\fill}}lclc}
\toprule
\multicolumn{2}{c}{\textbf{Detection}} &
\multicolumn{2}{c}{\textbf{Timing*}} \\
\midrule
Precision   & 79.2\% & Median error       & 4.2\,ms \\
Recall      & 96.6\% & $\pm$20\,ms        & 70.2\% \\
Specificity & 76.0\% & $\pm$40\,ms        & 98.3\% \\
F1-score    & 87.0\% &                   &        \\
\bottomrule
\end{tabular*}
\begin{tablenotes}[flushleft]
\footnotesize
\item *Median absolute timing error over true positives.
\end{tablenotes}
\end{threeparttable}
\end{table}

\section{Experimental Results on Real-World Disturbance Data}\label{sec:results}

The proposed framework was applied to the full RTE dataset. Because the dataset does not provide localized ground-truth inception timestamps, performance is primarily assessed through manual verification on a subset of events.

\subsection{Manual Evaluation}

Performance was assessed on a manually reviewed subset ($N=300$) constructed through stratified sampling across detector branches and rejected cases (Section~\ref{sec:eval_prot}). The detector achieved high recall (96.6\%) with moderate precision (79.2\%), reflecting a conservative operating point that prioritizes missed-fault avoidance (Table~\ref{tab:expert_overall}). False positives primarily correspond to switching or transient events with fault-like signatures.

Timing accuracy was evaluated relative to the reviewer-marked reference onset. The median absolute timing error is \SI{4.2}{ms}, with most detections within $\pm\SI{20}{ms}$ (70.2\%) and nearly all within $\pm\SI{40}{ms}$ (98.3\%), indicating sub-cycle localization performance. A simple $dI/dt$ baseline achieves lower recall (39.7\%) despite accurate localization when triggered. A branch-wise breakdown (Table~\ref{tab:expert_branches}) shows that the fast-path and high-confidence fusion branches are most reliable, whereas mid-confidence fusion accounts for most false positives.

\begin{table}[t]
\caption{Manual evaluation results by detector branch on the reviewed subset.}
\label{tab:expert_branches}
\centering
\footnotesize
\begin{threeparttable}
\begin{tabular}{lccccc}
\toprule
\textbf{Branch} & \textbf{N} & \textbf{TP} & \textbf{FP} & \textbf{Precision} & \textbf{Median timing error*} \\
\midrule
Fast path       & 80 & 71 & 9  & 88.8\% & 2.5\,ms  \\
Fusion high     & 40 & 35 & 5  & 87.5\% & 11.6\,ms \\
Fusion mid      & 33 & 14 & 19 & 42.4\% & 32.0\,ms \\
Fallback        & 7  & 6  & 1  & 85.7\% & 48.6\,ms \\
\bottomrule
\end{tabular}
\begin{tablenotes}[flushleft]
\footnotesize
\item *Timing error is the median absolute fault inception error over TP only.
\end{tablenotes}
\end{threeparttable}
\end{table}

\subsection{Dataset-Scale Behavior}

Across the dataset, the detector produced onset estimates for 5{,}959 of 12{,}053 recordings (49.4\%). This reflects the heterogeneous nature of the RTE data, which includes switching and other non-fault disturbances due to waveform-based preselection. Among detected events, negative-sequence evidence was present in all cases and zero-sequence evidence in 85.1\%, while the fast-path activated in 83.1\%. The mean estimated inception time was $\mu_{t_0}=\SI{0.258}{s}$ with standard deviation $\sigma_{t_0}=\SI{0.254}{s}$, characterizing typical detector behavior.

\section{Conclusion}

This paper presented a physics-guided framework for fault inception detection in real-world transmission disturbance data. Evaluated on 12{,}053 RTE recordings with manual validation, the method achieves high sensitivity and sub-cycle median timing accuracy. Robust normalization and physics-guided fusion enable reliable onset localization despite switching transients, saturation, and noise. Although demonstrated on 90\,kV data, the approach is largely voltage-level independent and can be adapted to other networks and sampling rates with minor parameter adjustments.

Observed failure cases mainly involve disturbances that mimic fault signatures, such as switching operations or transient resonance, which can produce temporary imbalance or current spikes. Conversely, slowly developing disturbances may not generate sufficient trigger evidence and can remain undetected. Beyond detection itself, the generated inception timestamps provide structured annotations for protection-oriented analysis of real-world disturbance data. Such annotations can support disturbance screening, relay-performance analysis, protection-oriented post-event review, and the construction of labeled datasets for data-driven methods. Future work will explore the use of these automatically generated labels for machine-learning-based disturbance classification and other data-driven protection applications using large-scale real-world disturbance data.

\section*{Acknowledgment}
This project was funded by the Deutsche Forschungsgemeinschaft (DFG, German Research Foundation) - 535389056.

\bibliographystyle{IEEEtran}
\bibliography{references}

\end{document}

%% file: acronyms.tex
\newacro{ai}[AI]{artificial intelligence}
\newacro{cnn}[CNN]{convolutional neural network}
\newacro{dfr}[DFR]{digital fault recorder}
\newacro{dl}[DL]{deep learning}
\newacro{hht}[HHT]{hilbert-huang transform}
\newacro{lstm}[LSTM]{long short-term memory}
\newacro{mad}[MAD]{median absolute deviation}
\newacro{wt}[WT]{wavelet transform}

%% file: references.bib
@misc{presvots_database_2024,
	title = {Database of {Voltage} and {Current} {Samples} {Values} from the {French} {Electricity} {Transmission} {Grid}, {Réseau} de {Transport} d'{Electricité} ({RTE}), {France}},
	url = {https://dfrdb.rte-france.com/},
	publisher = {https://github.com/rte-france/digital-fault-recording-database/},
	author = {Presvôts, Corentin and Prevost, Thibault},
	year = {2024},
}

@inproceedings{oelhaf_unsupervised_2025,
	address = {Atlanta, GA, USA},
	title = {Unsupervised {Clustering} for {Fault} {Analysis} in {High}-{Voltage} {Power} {Systems} {Using} {Voltage} and {Current} {Signals}},
	volume = {27},
	copyright = {Creative Commons Attribution 4.0 International},
	url = {https://arxiv.org/abs/2505.17763},
	doi = {10.48550/ARXIV.2505.17763},
	urldate = {2025-06-12},
	booktitle = {Fault and {Disturbance} {Analysis} {Conference} 2025},
	publisher = {arXiv},
	author = {Oelhaf, Julian and Kordowich, Georg and Maier, Andreas and Jager, Johann and Bayer, Siming},
	month = jun,
	year = {2025},
}

@article{boyd_learning_2024,
	title = {Learning from {Power} {Signals}: {An} {Automated} {Approach} to {Electrical} {Disturbance} {Identification} within a {Power} {Transmission} {System}},
	volume = {24},
	issn = {1424-8220},
	shorttitle = {Learning from {Power} {Signals}},
	url = {https://www.mdpi.com/1424-8220/24/2/483},
	doi = {10.3390/s24020483},
	language = {en},
	number = {2},
	urldate = {2026-03-13},
	journal = {Sensors},
	author = {Boyd, Jonathan D. and Tyler, Joshua H. and Murphy, Anthony M. and Reising, Donald R.},
	month = jan,
	year = {2024},
	pages = {483},
}

@book{phadke_computer_2009,
	edition = {1},
	title = {Computer {Relaying} for {Power} {Systems}},
	copyright = {http://doi.wiley.com/10.1002/tdm\_license\_1.1},
	isbn = {978-0-470-05713-1 978-0-470-74972-2},
	url = {https://onlinelibrary.wiley.com/doi/book/10.1002/9780470749722},
	doi = {10.1002/9780470749722},
	language = {en},
	urldate = {2026-03-13},
	publisher = {Wiley},
	author = {Phadke, Arun G. and Thorp, James S.},
	month = jul,
	year = {2009},
}

@article{zhang_transmission_2007,
	title = {Transmission {Line} {Boundary} {Protection} {Using} {Wavelet} {Transform} and {Neural} {Network}},
	volume = {22},
	copyright = {https://ieeexplore.ieee.org/Xplorehelp/downloads/license-information/IEEE.html},
	issn = {0885-8977},
	url = {http://ieeexplore.ieee.org/document/4141144/},
	doi = {10.1109/TPWRD.2007.893596},
	number = {2},
	urldate = {2026-03-13},
	journal = {IEEE Transactions on Power Delivery},
	author = {Zhang, Nan and Kezunovic, Mladen},
	month = apr,
	year = {2007},
	pages = {859--869},
}

@article{leys_detecting_2013,
	title = {Detecting outliers: {Do} not use standard deviation around the mean, use absolute deviation around the median},
	volume = {49},
	copyright = {https://www.elsevier.com/tdm/userlicense/1.0/},
	issn = {00221031},
	shorttitle = {Detecting outliers},
	url = {https://linkinghub.elsevier.com/retrieve/pii/S0022103113000668},
	doi = {10.1016/j.jesp.2013.03.013},
	language = {en},
	number = {4},
	urldate = {2026-03-13},
	journal = {Journal of Experimental Social Psychology},
	author = {Leys, Christophe and Ley, Christophe and Klein, Olivier and Bernard, Philippe and Licata, Laurent},
	month = jul,
	year = {2013},
	pages = {764--766},
}

@article{rousseeuw_robust_2011,
	title = {Robust statistics for outlier detection},
	volume = {1},
	copyright = {http://onlinelibrary.wiley.com/termsAndConditions\#vor},
	issn = {1942-4787, 1942-4795},
	url = {https://wires.onlinelibrary.wiley.com/doi/10.1002/widm.2},
	doi = {10.1002/widm.2},
	language = {en},
	number = {1},
	urldate = {2026-03-13},
	journal = {WIREs Data Mining and Knowledge Discovery},
	author = {Rousseeuw, Peter J. and Hubert, Mia},
	month = jan,
	year = {2011},
	pages = {73--79},
}

@inproceedings{perez_guide_2010,
	address = {College Station, TX, USA},
	title = {A guide to digital fault recording event analysis},
	isbn = {978-1-4244-6073-1},
	url = {http://ieeexplore.ieee.org/document/5469501/},
	doi = {10.1109/CPRE.2010.5469501},
	urldate = {2026-03-13},
	booktitle = {2010 63rd {Annual} {Conference} for {Protective} {Relay} {Engineers}},
	publisher = {IEEE},
	author = {Perez, Joe},
	month = mar,
	year = {2010},
	pages = {1--17},
}

@article{raichura_2_tddifftotal_2021,
	title = {[2\_TD\${DIFF}]{Total} {Harmonic} {Distortion} ({THD}) based discrimination of normal, inrush and fault conditions in power transformer},
	volume = {36},
	issn = {17550084},
	url = {https://linkinghub.elsevier.com/retrieve/pii/S1755008420300661},
	doi = {10.1016/j.ref.2020.12.001},
	language = {en},
	urldate = {2026-03-13},
	journal = {Renewable Energy Focus},
	author = {Raichura, Maulik and Chothani, Nilesh and Patel, Dharmesh and Mistry, Khyati},
	month = mar,
	year = {2021},
	pages = {43--55},
}

@inproceedings{wang_analysis_2008,
	address = {College Station, TX, USA},
	title = {Analysis of {Transformer} {Inrush} {Current} and {Comparison} of {Harmonic} {Restraint} {Methods} in {Transformer} {Protection}},
	isbn = {978-1-4244-1949-4},
	url = {http://ieeexplore.ieee.org/document/4515052/},
	doi = {10.1109/CPRE.2008.4515052},
	urldate = {2026-03-13},
	booktitle = {2008 61st {Annual} {Conference} for {Protective} {Relay} {Engineers}},
	publisher = {IEEE},
	author = {Wang, Jialong and Hamilton, Randy},
	month = apr,
	year = {2008},
	pages = {142--169},
}

@article{kezunovic_smart_2011,
	title = {Smart {Fault} {Location} for {Smart} {Grids}},
	volume = {2},
	copyright = {https://ieeexplore.ieee.org/Xplorehelp/downloads/license-information/IEEE.html},
	issn = {1949-3053, 1949-3061},
	url = {http://ieeexplore.ieee.org/document/5723781/},
	doi = {10.1109/TSG.2011.2118774},
	number = {1},
	urldate = {2026-03-13},
	journal = {IEEE Transactions on Smart Grid},
	author = {Kezunovic, Mladen},
	month = mar,
	year = {2011},
	pages = {11--22},
}

@article{moreto_using_2015,
	title = {Using disturbance records to automate the diagnosis of faults and operational procedures in power generators},
	volume = {9},
	issn = {1751-8687, 1751-8695},
	url = {https://ietresearch.onlinelibrary.wiley.com/doi/10.1049/iet-gtd.2014.0785},
	doi = {10.1049/iet-gtd.2014.0785},
	language = {en},
	number = {15},
	urldate = {2026-03-13},
	journal = {IET Generation, Transmission \& Distribution},
	author = {Moreto, Miguel and Rolim, Jacqueline G.},
	month = nov,
	year = {2015},
	pages = {2389--2397},
}

@article{rodrigues_power_2023,
	title = {Power {Quality} {Transient} {Detection} and {Characterization} {Using} {Deep} {Learning} {Techniques}},
	volume = {16},
	issn = {1996-1073},
	url = {https://www.mdpi.com/1996-1073/16/4/1915},
	doi = {10.3390/en16041915},
	language = {en},
	number = {4},
	urldate = {2026-03-17},
	journal = {Energies},
	author = {Rodrigues, Nuno M. and Janeiro, Fernando M. and Ramos, Pedro M.},
	month = feb,
	year = {2023},
	pages = {1915},
}

@inproceedings{hossack_multiagent_2003,
	address = {Toronto, Ont., Canada},
	title = {A multiagent architecture for protection engineering diagnostic assistance},
	isbn = {978-0-7803-7989-3},
	url = {http://ieeexplore.ieee.org/document/1270379/},
	doi = {10.1109/PES.2003.1270379},
	urldate = {2026-03-17},
	booktitle = {2003 {IEEE} {Power} {Engineering} {Society} {General} {Meeting} ({IEEE} {Cat}. {No}.{03CH37491})},
	publisher = {IEEE},
	author = {Hossack, J.A. and Menal, J. and McArthur, S.D.J. and McDonald, J.R.},
	year = {2003},
	pages = {640},
}

@article{mozaffari_real-time_2022,
	title = {Real-{Time} {Detection} and {Classification} of {Power} {Quality} {Disturbances}},
	volume = {22},
	issn = {1424-8220},
	url = {https://www.mdpi.com/1424-8220/22/20/7958},
	doi = {10.3390/s22207958},
	language = {en},
	number = {20},
	urldate = {2026-03-17},
	journal = {Sensors},
	author = {Mozaffari, Mahsa and Doshi, Keval and Yilmaz, Yasin},
	month = oct,
	year = {2022},
	pages = {7958},
}
